\RequirePackage[2020-02-02]{latexrelease}
\documentclass[prl,showpacs,preprintnumbers,amsmath,amssymb]{revtex4}
\usepackage[dvips]{graphics}
\usepackage{tikz}

\begin{document}

\title{Matrix Mechanics of a Particle in a One-Dimensional Infinite Square Well}

\author{Vlatko Vedral}
\affiliation{Clarendon Laboratory, University of Oxford, Parks Road, Oxford OX1 3PU, United Kingdom}

\date{\today}

\begin{abstract}
\noindent We solve the infinite potential well problem using the methods of Heisenberg's matrix mechanics. In addition to being of educational value, the matrix mechanics allows us to deal with various unphysical issues caused by this potential in a seemingly unproblematic fashion. We also show how to treat many particles within this representation. 
\end{abstract}


\maketitle                           

A particles confined in an  infinite potential square well is possibly the first probelm encoutered in any quantum mechanical textbook. It is almost always solved in the Schr\"odinger picture by writing down the time-independent Schr\"odinger equation and thereby obtaining the particle's energy eigenstates and eigenvalues. Let us say that the infinite walls are located at $x=0$ and $x=L$. Then, the normalized $n$th eigenstate is
\begin{equation}
\Psi_n (x) = \sqrt{\frac{2}{L}}\sin k_n x
\end{equation}
where $k_n = n\pi/L$. Its corresponding eigenvalue is 
\begin{equation}
E_n = \frac{\hbar^2 k^2_n}{2m} = \frac{\hbar^2 \pi^2 n^2}{2mL^2}
\end{equation}
where $m$ is the mass of the particle. The eigenstates clearly satisfy the boundary condition, namely that the wavefunction vanishes at $x=0$ and $x=L$, $\Psi_n (0)=\Psi_n (L)=0$. The time evolution simply adds an extra phase factor $e^{-i\omega_n t}$ to the $n$th eigenstate, where $\omega_n = \hbar \pi^2 n^2/2mL^2$.

Imagine now that we would like to find the expected value of $V$ or $dV/dx$ in order to phrase the Ehrenfest theorem (which tells us that on average, $\langle dp/dt \rangle = \langle dV/dx \rangle$, i.e. quantum mechanics reproduces the classical behaviour as captured by Newton's second law $ma = F$). However, the problem with calculating $\langle V \rangle$ is that any average of the kind $\int \psi_n V \psi_m dx$ seems ill defined \cite{Rokhsar}. This is because the domain where $V$ is non-zero is exactly orthogonal to the domain where the eigenstates are non-zero other than at $x=0,L$ where the potential is infinite and the states are zero. Here the product of the two, $\infty \times 0$, seems undefined.  One might be tempted to think that Ehrenfest's theorem does not apply here and that there is no well defined classical limit in the case of the infinite square well. 

This, however, would be the wrong conclusion, not least because it would contradict Bohr's correspondence principle. This principle stipulates that quantum physics must - in some limit - recover the classical behaviour \cite{Longair}. To see how the matrix elements of the gradient of the potential can be calculated, we 
now express this problem in the Heisenberg matrix formulation \cite{Born}. Because in the Heisenberg formulation the equations of motion are simply the classical equations of Hamilton, the correspondence principle is automatically satisfied \cite{Green,Werden}. If $x$ and $p$ are particle's position and momentum respectively, and $H$ is the Hamiltonian, we have that 
\begin{equation} 
\frac{\partial x}{\partial t} = \frac{\partial H}{\partial p} \;\;\;\;\; \frac{\partial p}{\partial t} =- \frac{\partial H}{\partial x}
\end{equation}
The difference between classical and quantum mechanics lies entirely in the fact that in quantum mechanics $x$ and $p$ are matrices. Note that 
the derivative of the $H$ operator with respect to the $p$ operator is defined as 
\begin{equation} 
\frac{\partial H}{\partial p_x} = \lim_{\epsilon \rightarrow 0} \frac{H(...,p+\epsilon I,...)-H(...,p,...)}{\epsilon}
\end{equation}
and likewise for the derivative with respect to $x$. 

The key, therefore, is to find the matrices representing the position and the momentum of the particle in the basis that diagonalises the Hamiltonian. The position matrix elements are defined as
\begin{eqnarray}
x_{kl} = \int x \Psi_k (x)\Psi_l (x) dx 
\end{eqnarray}
and, using the above eigenstates, we obtain the off-diagonal position matrix elements:
\begin{eqnarray}
x_{k\neq l} = -\frac{L}{\pi^2}\frac{\{(k-l)^2\cos ((k+l)\pi) - (k+l)^2\cos ((k-l)\pi)+4kl\}}{(k^2-l^2)^2}\; .
\end{eqnarray}
The diagonal elements are $x_{nn} = L/2$ for all $n$. The temporal dependence is given by 
\begin{equation}
x_{kl}(t) = x_{kl} e^{-i(\omega_k - \omega_l)t} \; .
\end{equation}
The momentum matrix is then given by 
\begin{eqnarray}
p_{kl} = m \frac{d x_{kl}}{dt} =  \frac{i\hbar}{2L}\frac{\{(k-l)^2\cos ((k+l)\pi) - (k+l)^2\cos ((k-l)\pi)+4kl\}}{(k^2-l^2)}
\end{eqnarray}
with all of the diagonal elements vanishing (also derived in \cite{Prentis}). One can check that $p^2$ is a diagonal matrix whose $n$th element is $(p^2)_{nn}=n^2\pi^2/L^2$ meaning that the Hamiltonian is automatically diagonalised. Also, it is straightforward to verify that the matrices $x$ and $p$
satisfy the Heisenberg commutation relations $xp-px = i\hbar I$, where $I$ is the identity matrix. 

The matrix elements for the force $dV/dx$ can be found from the momentum equation of motion:
\begin{eqnarray}
\frac{dp_{kl}}{dt} =  \bigg(\frac{dV}{dx}\bigg)_{kl} \; .
\end{eqnarray}
From this, the Ehrenfest theorem is an immediate consequence, since the average of the above equiality in an arbitrary state $\Psi = \sum_n a_n \Psi_n$ leads to 
\begin{eqnarray}
\sum_{kl}a_k^*a_l\frac{dp_{kl}}{dt} = m \sum_{kl}a_k^*a_l \bigg(\frac{dV}{dx}\bigg)_{kl} \; .
\end{eqnarray}
Therefore, a result that seems difficult to obtain in the Schr\"odinger picture follows more directly in the Heisenberg picture. In fact, as we said, it is built into it from the start in the Heisenberg picture.

It is now instructive to discuss the spreading of the wave-packet confined to such an infinite square well potential. We wish to compute the commutator $[x(t),x(0)] = x(t) x(0) - x(0)x(t)$ since this is a lower bound on the product of the dispersions $\Delta x(t) \Delta x(0)$. 

This could be used to explain the collapses and revivals of wave-packets in the infinite potential well. For very short time intervals $\delta t$ (compared to $Lm/p$), the behaviour of the wave-packet in the well is the same as that for a free particle. This is because the only place where the wave-packet experiences force is at the walls. Far away from the walls, we have that 
\begin{equation}
x(\delta t) = x(0)+\frac{p}{m} \delta t
\end{equation} 
which leads us to conclude that \cite{Sch}
\begin{equation}
\Delta x(\delta t) \Delta x(0)\geq \frac{\hbar\delta t}{2m}
\end{equation} 
As the wave-packet continues to spread, it ultimately becomes flatter and this is known as the collapse of the wave-packet.

However, for longer times it is clear that the force term $\bigg(\frac{dV}{dx}\bigg)_{kl}$ will contribute to this. Namely, the equation of motion will be 
\begin{equation}
x(\delta t) = x(0)+\frac{p}{m} \delta t - \frac{1}{2m}\frac{dV}{dx} (\delta t)^2
\end{equation} 
So much so, in fact, that the wave-packet will start to shrink and in fact undergo a revival of its position spread. 
This is easily seen from the fact that at the time $t_r = 4mL^2/\hbar\pi$, we have that $x(t_r)=x(0)$. This is because all the phase factors $e^{-i(\omega_n-\omega_m)t_r} =1$. Therefore, $\Delta x(t_r) = \Delta x(0)$, clealry signalling the revival of the wave-packet.

A great deal of research has been done in this direction, but the main point of our work has been to show the advantages of the Heisenberg picture. We can avoid manipulations of non-differentiable functions such as the delta function (sometimes referred to as a generalised function), the step function (whose derivative is the delta function) and all the resulting potentially pathological behaviour associated with them. 

Finally, it is simple to treat more than one particle in the well. The easiest is to ``second quantise" which means upgrading the wave-function $\Psi_n (x)$ into an operator. It then becomes
\begin{equation}
\hat\Psi (x) = \sum_n \sqrt{\frac{2}{L}}\hat a_n \sin k_n x
\end{equation}
where now the amplitude of the wave also becomes an operator $\hat a_n$. The physical meaning is that the 
operator $\hat\Psi (x)$ annihilates a particle at position $x$, while $\hat a_n$ annihilates the particle in the whole mode $n$ (the mode here is just a synonym for an energy eigenstate). Fermionic or bosonic nature of the particles is included by imposing the (anti)-commutation relations on the creation and annihilation operators:
\begin{equation}
[\hat\Psi (x), \hat \Psi^\dagger (x')]_{\pm} = \delta (x-x') \; ,
\end{equation}
where the plus/minus signs correspond to fermions/bosons respectively. Equivalently we have that $[\hat a_n,\hat a^\dagger_m]_{\pm} = \delta_{nm}$. In other words, operators pertaining to different modes always (anti)-commute.

The Hamiltonian now becomes:
\begin{equation}
\hat H = \sum_n \hbar \omega_n \hat a^\dagger_n \hat a_n ; .
\end{equation}
This operator has the usual meaning in that it counts the number of particles in each level, multiplies it by the corresponding energy of that level, and then sums up over all energy levels. 

In the Heisenberg picture, the field operators evolve in time, while the Heisenberg state is stationary. For instance, if we have a condensate of $N$ bosonic atoms all in the lowest energy level the state is given by
\begin{equation}
|\Psi\rangle = \frac{(a^\dagger_1)^N}{\sqrt{N!}} |0\rangle \; .
\end{equation}
The operators then evolve in the usual way as $\hat \Psi (x,t) = e^{i\hat H t} \hat \Psi (x) e^{-i\hat H t}$.
In that way, all physically meaningful quantities are obtained by taking the expectation values of the kind $\langle \Psi| \hat O(x,t)|\Psi\rangle$. The expected number of particles at the position $x$ and time $t$ is, for instance, given by: $\langle \Psi| \hat \Psi^\dagger (x,t)\hat \Psi (x,t)|\Psi\rangle$.  All other features, such as the internal degrees of freedom or interactions between particles can be included in the usual manner done in quantum field theory \cite{Weinberg}.

\textit{Acknowledgments}: VV's research is supported by the Moore and Templeton Foundations.

\end{document}